\documentclass[%
preprint,
superscriptaddress,
 amsmath,amssymb,
 aps,
floatfix,
]{revtex4-1}
\usepackage{graphicx}
\usepackage{dcolumn}
\usepackage{bm}
\usepackage{mathrsfs}
\usepackage{natbib}
\usepackage[normalem]{ulem}
\usepackage[dvipsnames]{xcolor}
\usepackage{hyperref}

\begin{document}

\title{Ghost-imaging-enhanced non-invasive spectral characterization of stochastic x-ray free-electron-laser pulses} 
\author{Kai Li}\email{kail@anl.gov}
 \affiliation{Department of Physics, The University of Chicago, Chicago, IL 60637 USA}
 \affiliation{Chemical Sciences and Engineering Division, Argonne National Laboratory, Lemont, IL 60439 USA}
  

\author{Joakim Laksman}
\affiliation{European XFEL, Holzkoppel 4, 22869 Schenefeld 
Germany}

\author{Tommaso Mazza}
\affiliation{European XFEL, Holzkoppel 4, 22869 Schenefeld 
Germany}

\author{Gilles Doumy}
\affiliation{Chemical Sciences and Engineering Division, Argonne National Laboratory, Lemont, IL 60439 USA}

\author{Dimitris Koulentianos}
\affiliation{Chemical Sciences and Engineering Division, Argonne National Laboratory, Lemont, IL 60439 USA}

\author{Alessandra Picchiotti}
\affiliation{The Hamburg Centre for Ultrafast Imaging, Hamburg University, Luruper Chaussee 149, 22761, Hamburg, Germany}

\author{Svitozar Serkez}
\affiliation{European XFEL, Holzkoppel 4, 22869 Schenefeld 
Germany}

\author{Nina Rohringer}
\affiliation{Center for Free-Electron Laser Science CFEL, Deutsches Elektronen-Synchrotron DESY, Notkestra{\ss}e 85, 22607 Hamburg, Germany}
\affiliation{Department of Physics, Universität Hamburg, 20355 Hamburg, Germany}

\author{Markus Ilchen}
\affiliation{Deutsches Elektronen-Synchrotron DESY, Notkestra{\ss}e 85, 22607 Hamburg, Germany}

\author {Michael Meyer}
\affiliation{European XFEL, Holzkoppel 4, 22869 Schenefeld 
Germany}

\author{Linda Young}
 \email{young@anl.gov}
 \affiliation{Department of Physics, The University of Chicago, Chicago, IL 60637 USA}
 \affiliation{Chemical Sciences and Engineering Division, Argonne National Laboratory, Lemont, IL 60439 USA}
 \affiliation{James Franck Institute, The University of Chicago, Chicago, IL 60637 USA}

\date{\today}

\begin{abstract}
High-intensity ultrashort X-ray free-electron laser (XFEL) pulses are revolutionizing the study of fundamental nonlinear x-ray matter interactions and coupled electronic and nuclear dynamics. To fully exploit the potential of this powerful tool for advanced x-ray spectroscopies, a noninvasive spectral characterization of incident stochastic XFEL pulses with high resolution is a key requirement. Here we present a methodology that combines high-acceptance angle-resolved photoelectron time-of-flight spectroscopy and ghost imaging to enhance the quality of spectral characterization of x-ray free-electron laser pulses. Implementation of this non-invasive high-resolution x-ray diagnostic can greatly benefit the ultrafast x-ray spectroscopy community by functioning as a transparent beamsplitter for applications such as transient absorption spectroscopy in averaging mode as well as covariance-based x-ray nonlinear spectroscopies in single-shot mode where the shot-to-shot fluctuations inherent to a self-amplified spontaneous emission (SASE) XFEL pulse are a powerful asset.
    
\end{abstract}

\maketitle

\clearpage
\section*{Introduction}

X-ray free-electron lasers, with brilliance ten orders of magnitude higher than synchrotrons, continuous tunability over the soft and hard x-ray regimes and sub-femtosecond pulse duration \cite{duris2020tunable}, have emerged as a powerful tool both to explore fundamental nonlinear x-ray interactions in isolated atomic and molecular systems \cite{young2010femtosecond,Hoener-2010-PRL,Doumy-2011-PRL,Kanter-2011-PRL,Rudenko-2017-Nature,Mazza-2020-PRX}, and, to follow photoinduced electronic and nuclear dynamics on their intrinsic femtosecond timescales via pump/probe techniques \cite{young2018roadmap,Wolf-2017-NatComm}. For the latter objective, core-level x-ray transient absorption (XTAS) with ultrafast x-ray pulses has become a workhorse - it projects core electronic states onto unoccupied valence/Rydberg states, thereby capturing the evolution of valence electronic motion following an excitation pulse. However, realization of XTAS is challenging at x-ray free-electron lasers (XFELs) where the x-ray pulses with bandwidth $\Delta E/E \sim 1\%$, typically produced by self-amplified spontaneous emission (SASE), have spiky temporal and spectral profiles that vary stochastically on a shot-by-shot basis  \cite{kim1986analysis,andruszkow2000first,milton2001exponential,Geloni-2010-NJP,hartmann2018attosecond}. The traditional approach for XTAS with XFELs is to monochromatize the SASE beam \cite{Lemke-2013-JPCA,Higley-2016-RSI} and scan the monochromatic beam ($\Delta E/E\sim 0.01\%$) across the desired spectral range. This makes inefficent use of the full XFEL beam, imposes limits on time resolution via the uncertainty principle, and, by reducing the pulse intensities, hampers realization of nonlinear x-ray spectroscopies. An alternative approach is to monitor incident and transmitted intensity to obtain an absorption spectrum, $I_T(\omega)/I_0(\omega)$, across the entire SASE bandwidth. With this approach one may realize experimental techniques employing correlation analysis that take advantage of the intrinsic stochastic nature of XFELs pulses \cite{Kimberg2016SD,ratner2019pump,cavaletto2021high,kayser2019core}. By using pulses with uncorrelated fluctuations one can leverage the noise such that each repetition of the experiment, i.e. each XFEL shot, represents a new measurement under different conditions.  As an example, spectral ghost imaging has been applied to obtain an absorption spectrum with energy resolution better than the averaged SASE bandwidth \cite{driver2020attosecond,li2021time}. In general, the characterization of the incident pulses is essential to this class of covariance spectroscopies as previously demonstrated in the UV regime \cite{Tollerud2019PNAS}.

Several diagnostic tools have demonstrated well-resolved spectral measurements on a single-shot basis without compromising the quality of the x-ray beam. A commonality is the use of optical elements to split the incident x-ray beam into reference and sample beams. Beamsplitters for hard x-rays use crystal Bragg diffraction \cite{Zhu-2012-APL,makita2015high} while diffraction gratings are used for soft x-rays \cite{engel2020parallel,Brenner2019OE}. An alternative is to use photoionization of a dilute target gas and measure the kinetic energy of ejected photoelectrons to retrieve the incident photon spectrum via the photoelectric effect   \cite{viefhaus2013variable,laksman2019commissioning,Walter:ys5104}. Indeed, the use of an array of 16 electron time-of-flight spectrometers (eTOFs) radially distributed about the propagating x-ray beam and hereafter referred to as the photoelectron spectromater array, (PES array) has enabled the measurement of the position, polarization, and central energy of an x-ray photon beam as demonstrated at the PETRA-P04 beamline \cite{viefhaus2013variable}. At XFELs, while it is straightforward to measure the central photon energy with the PES array \cite{laksman2019commissioning} as has been demonstrated for two-color x-ray pulses \cite{serkez2020opportunities} and to obtain simultaneous polarization diagnostics  \cite{lutman2016polarization,hartmann2016circular}, it is more challenging to obtain single-shot spectra with an energy resolution comparable to a grating spectrometer. 

Here we use a ghost-imaging algorithm to improve the energy resolution of the raw PES array measurements. Thousands of SASE spectra were measured simultaneously by the PES array and a grating spectrometer and ghost imaging was applied to compute the response matrix of the PES array. The response matrix was then used to reconstruct the x-ray spectrum with energy resolution improved from $\sim1$ eV to 0.5 eV at a central energy of 910 eV for a resolution of $\Delta E/E \sim 1/2000$ under the present conditions.  This response matrix derived from ghost imaging also provides predictive power for the spectral profile of yet-to-be-measured XFEL pulses.

\section*{Results}
\subsection*{Spectral ghost imaging}
Ghost imaging is an experimental technique which uses statistical fluctuations of an incident beam to extract information about an object using a beam replica that has not physically interacted with the object \cite{Padgett2017PTRS}. It can be used in the spatial \cite{pelliccia2016experimental,yu2016fourier, kim2020ghost}, temporal \cite{ratner2019pump} and spectral \cite{driver2020attosecond,li2021time} domains. Traditional ghost imaging requires a beam splitter to separate the incident beam into two replicas, the object beam and the reference beam. The object beam interacts with the sample and a low-resolution detector is used to measure the signal whose intensity is proportional to the interaction and the incident beam. The reference beam is directly measured by a high-resolution detector to extract knowledge of the incident beam. The incident light source varies shot-by-shot and numerous measurements are carried out to calculate the correlation function between the two signals from the object and the reference beams. The correlation function of the measurements is analyzed to extract information of the sample. The advantage of ghost imaging is that the object beam does not necessarily need to be strong -- thus protecting the samples from radiation damage. In addition, due to the fluctuations of the light source and correlation analysis, ghost imaging is robust to noise and background signals.


Ghost imaging essentially maps the high-resolution signal onto the low-resolution one, making it an ideal tool to calibrate devices with high resolution. The correlation function generated by ghost imaging contains information on the response of a device to the different incident signals. This extracted information can be further used to correct defects or discrimination present in a device. The ghost imaging calibration method reconstructs a high quality signal that achieves resolution beyond the low-resolution instrumental limit. The stochastic nature of a SASE XFEL makes it well-suited for ghost imaging in the temporal and spectral domains. Here, ghost imaging is used to calibrate the eToFs of the PES array and obtain a response matrix, which is then applied to reconstruct a more accurate incident x-ray spectrum. One challenge for applying the ghost imaging method in the x-ray regime is the requirement of a beamsplitter. Although x-ray beamsplitters are available as mentioned above, the non-invasive gas-target measurement is suitable to replace the function of the beamsplitter.

\subsection*{Experimental procedure}
\begin{figure*}
\includegraphics[width=0.9\linewidth]{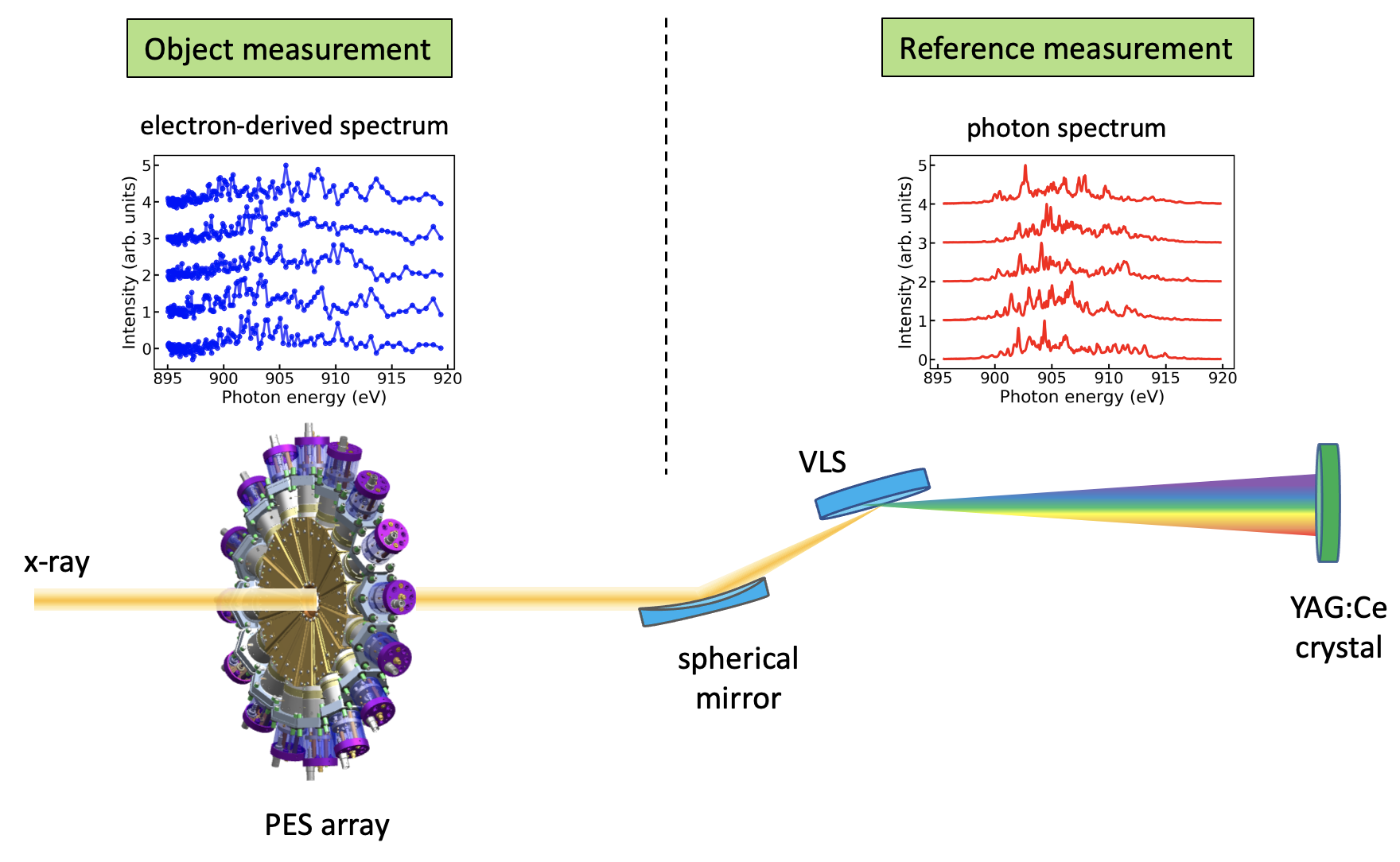}
\caption{\label{fig:setup} Schematic of the experimental layout and elements of ghost imaging. Self-amplified spontaneous emission (SASE) x-ray free-electron laser (XFEL) pulses first interact with dilute neon gas in the photoelectron spectrometer (PES) array where the kinetic energies of $1s$ photoelectrons are measured by the array of electron time-of-flight spectrometers (eToFs). These kinetic energies are used to produce the electron-derived spectrum that forms the object measurement. The transmitted x-ray pulse is then focused on the variable-line-spacing (VLS) grating by a spherical mirror and dispersed on a YAG:Ce crystal. The induced fluorescence is recorded by a charge-coupled device (CCD) as a 2D image from which we extract the single-shot reference measurement.}
\end{figure*}

The energy spectrum of the incident x-ray beam was characterized non-invasively by photoionization of dilute neon gas at the center of an array of 16-eToFs, i.e. the PES array \cite{laksman2019commissioning} as shown in Fig.~\ref{fig:setup}.  The arrival times of Ne $1s$ photoelectrons were measured by the eToFs located in the plane perpendicular to the beam propagation direction (see Supplementary Note 1). The ion and electron optics program SIMION was used to carry out trajectory simulations given the drift tube length and retardation voltages and thus establish a traditional calibration between the electron time-of-flight and kinetic energy, $E_k$. The incident photon energy was derived by adding the Ne $1s$ binding energy, 870 eV, to the measured $E_k$. The spectrum obtained by the PES array for several random shots using this traditional method is shown in Fig.~\ref{fig:setup} as the object measurement. Under the present experimental conditions, the energy resolution achievable by the PES array was around 1 eV, which is not comparable to the high-resolution grating spectrometer measurement where 0.2 eV FWHM ($\Delta E/E$) can be readily achieved.

After passing through the PES array, the same FEL beam was characterized by a spectrometer based on a VLS grating and a Ce:YAG screen as shown in Fig.~\ref{fig:setup} as the reference measurement. The PES array contains very dilute gas which does not attenuate or otherwise alter the x-ray beam. Thus, ideally the same spectrum would be obtained from the electron (PES array) and photon (grating spectrometer) measurements. However, the measurement of a single random shot shown in Fig.~\ref{fig:sigle1}\textbf{a} reveals differences. The grating spectrometer resolution is much higher than the resolution of the PES array, thus creating a large deviation between the two spectra. The use of ghost imaging to retrieve a response matrix which is then used to improve the performance of the PES array measurements is demonstrated in the following.

\subsection*{Principle of reconstruction}

Theoretically the photoelectron signal $c$ (after normalization to the gas density) is proportional to the incident photon spectrum $s$ as measured by the spectrometer 
\begin{equation}
c = A s
\label{gh1}
\end{equation}
where $A$ relates the PES array signals to the incident photon spectrum, is an $(m\times n)$ matrix with the PES array time-of-flight (ToF) points $m=137$ and the spectrograph pixels $n=1900$ in the region of interest between 895 and 920 eV. This equation resembles the basic equation in ghost imaging and is usually used to obtain sample information by solving for $A$. However, in order to predict the incident spectrum based on PES array measurements, we formally write equation (\ref{gh1}) as
\begin{equation}
s = R c
\label{gh2}
\end{equation}

where the response matrix $R$ is related to matrix $A$. $R$ maps the low-resolution PES array measurements to high-resolution grating spectrometer measurements. In other words, $R$ is a calibration matrix that contains information on the characteristics of the eToFs. After retrieving the response matrix $R$, according to equation (\ref{gh2}), it can be used to generate a high-resolution spectrum with the intrinsic defects and broadening of PES array removed.

\subsection*{Ghost imaging reconstructed spectrum}

To solve equation (\ref{gh2}), we take advantage of the $N$ independent measurements obtained. Each shot gives a realization of $s_i$ and $c_j$ in equation $s_i=\sum_{j=1}^{m}R_{ij} c_j$ with $m$ unknown variables $R_{ij}$. Combining all measurements gives $N$ independent linear equations which can be solved to uniquely determine the unknown variables if $N>m$. Instead of directly solving these equations, the response matrix elements are determined by least square regression, i.e. by minimizing the quantity $|s-Rc|^2$. Single-photon Ne $1s$ ionization exhibits a dipole angular distribution pattern due to the linear (horizontal) polarization of the x-rays and the spherical $1s$ electron orbital. To increase the signals, we combined six eToFs near the polarization direction which have strong 1s peaks, to form the PES array measurement vector $c$ with dimension $m=6\times137=822$. 

\begin{figure}
\centering
\includegraphics[width=0.5\linewidth]{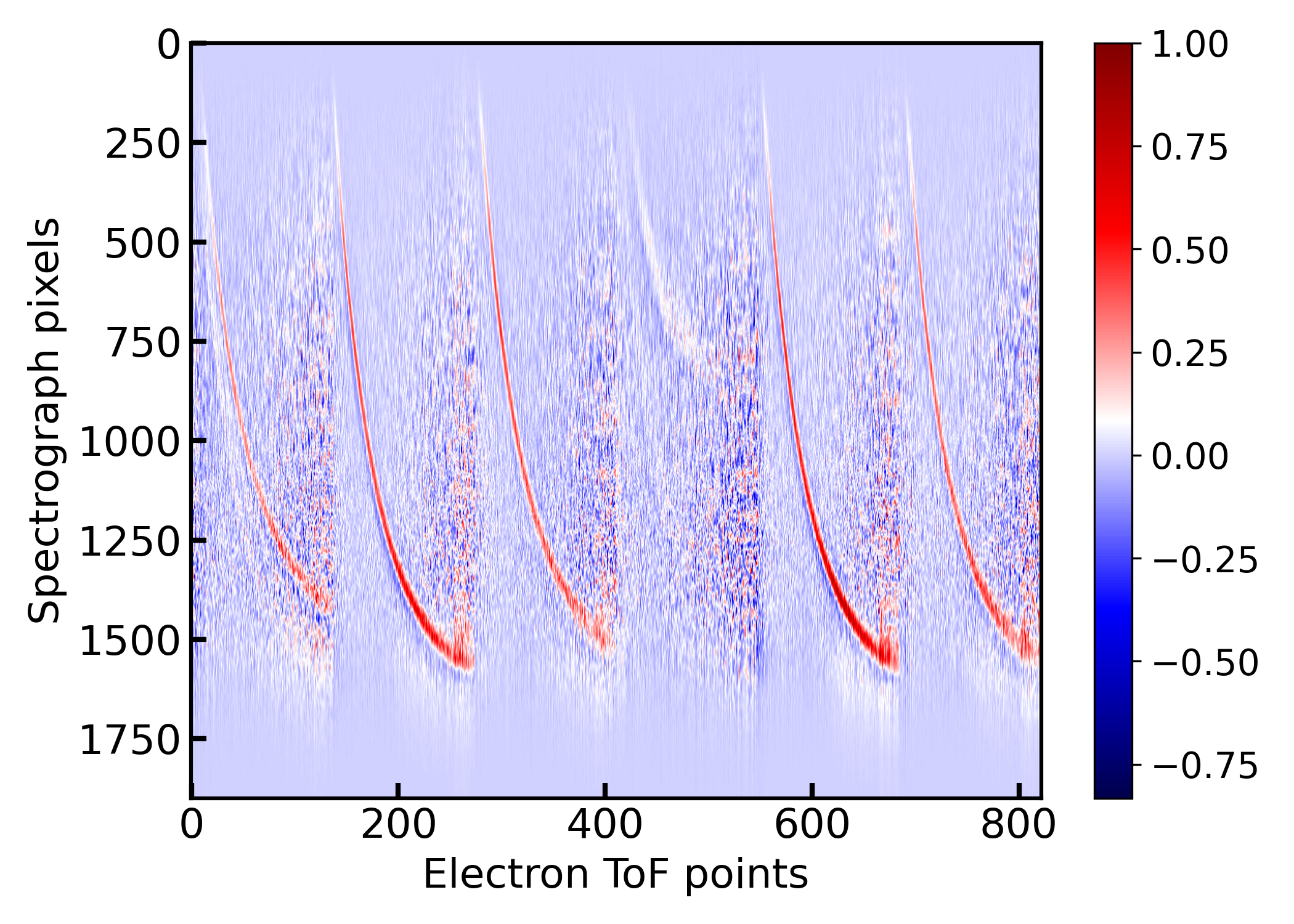}
\caption{\label{fig:R} The response matrix for the photoelectron spectrometer (PES) array. The response matrix, $R$ (Eqn. \ref{gh2}), was computed with data from six electron time-of-flight spectrometers (eToFs), the corresponding photon spectra from the variable line spacing (VLS) spectrometer and all the available shots $N=15337$. Each of the eToFs produces a distinct curve (red line) mapping the time-of-flight to the photon energy measured by the VLS spectrometer. The intensity at a specific photon energy, i.e. spectrograph pixel, is comprised of contributions (positive and negative) from each of the six eToFs.}
\end{figure}

The calculated PES array response matrix using all shots ($N =15337$) is shown in Fig.~\ref{fig:R}. Compared with the traditional calibration function, which just maps ToFs onto kinetic energy, here we retrieved a matrix whose values represent the sensitivity of the PES array to photons of different energy. As expected, there are six different calibration lines connecting the eToFs to spectrograph pixels. The linewidth represents the instrumental broadening. One eToF does not work well and gives relatively small signals. We tried different regression optimizers and got essentially the same response matrix, demonstrating the robustness of our method. As discussed below, the response matrix can be used to obtain a better spectrum. Note that one can quickly obtain the traditional calibration lines of eToFs, by fitting the lines within a nonconverged response matrix obtained by using only 1500 shots. 

\begin{figure}
\centering
\includegraphics[width=0.5\linewidth]{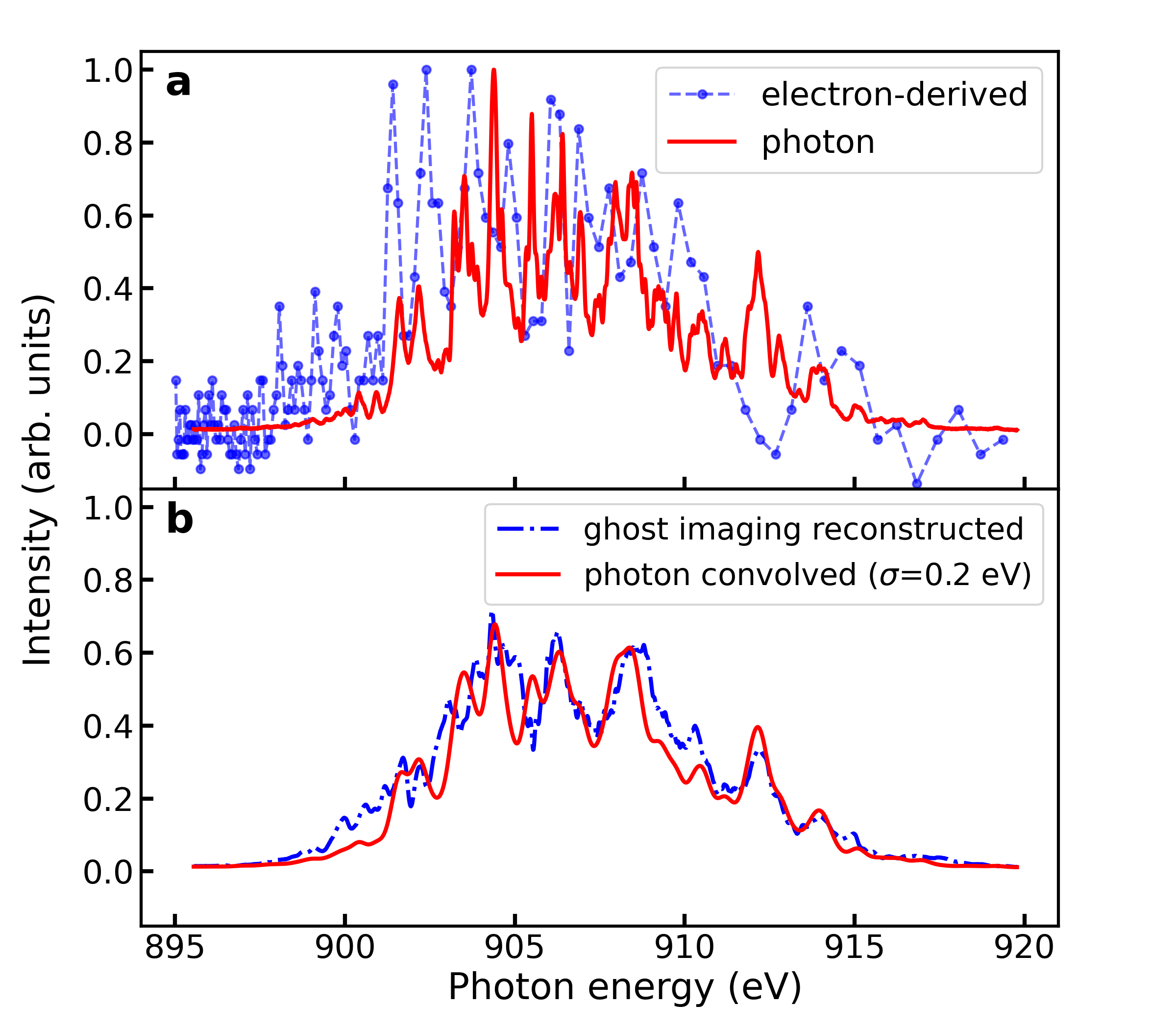}
\caption{\label{fig:sigle1} Improved single-shot spectral retrieval using the ghost-imaging algorithm. Single-shot electron spectra before (\textbf{a}) and after ghost-imaging reconstruction (\textbf{b}) are shown in blue compared to photon spectra in red. Top panel (\textbf{a}) shows the raw electron-based spectrum from a single eToF (blue dashed line) and grating-based photon spectrum (red solid line). Lower panel (\textbf{b}) shows the ghost-imaging-reconstructed electron spectrum (blue dot-dashed line) and the photon spectrum after convolution with a Gaussian ($e^{-x^2/(2\sigma^2)}$ with $\sigma=0.2$ eV (red solid line).}
\end{figure}

It is reasonable to assume the response matrix of the PES array does not change for a given photon energy, gas target, photoelectron energy range and PES array configuration (fixed retardation, bias voltage...); thus $R$ can be used to predict the spectra of new shots. Higher resolution electron spectra $s_r$ can be reconstructed by multiplying the response matrix $R$ by the PES array measurement $c$ according to equation (\ref{gh2}). As illustrated in Fig.~\ref{fig:sigle1}, the peak profile and intensities of the PES array data (\textbf{a}) are changed after multiplying the matrix with the PES array measurement (\textbf{b}).

The photon spectrum in Fig.~\ref{fig:sigle1} \textbf{b} was convolved with a Gaussian function with $\sigma$=0.2 eV to compare with the ghost-imaging-reconstructed spectrum. The $\sigma$ was derived by considering the number of data points in the PES array versus the spectrometer measurement. There are $1900/137\approx14$ spectrograph pixels between two eToF points; multiplying by the 0.013 eV/pixel dispersion of the spectrometer gives 0.2 eV which we take to be $\sigma$. One function of reconstruction is to remove instrumental broadening. Thus one observes the higher resolution of the reconstructed spectrum, which matches well with the convolved grating spectrometer measurement. This also indicates that in our case the resolution of the reconstructed spectrum is limited by the number of data points within the Ne 1s photoelectron peak of PES array signal.

\begin{figure}
\centering
\includegraphics[width=0.5\textwidth]{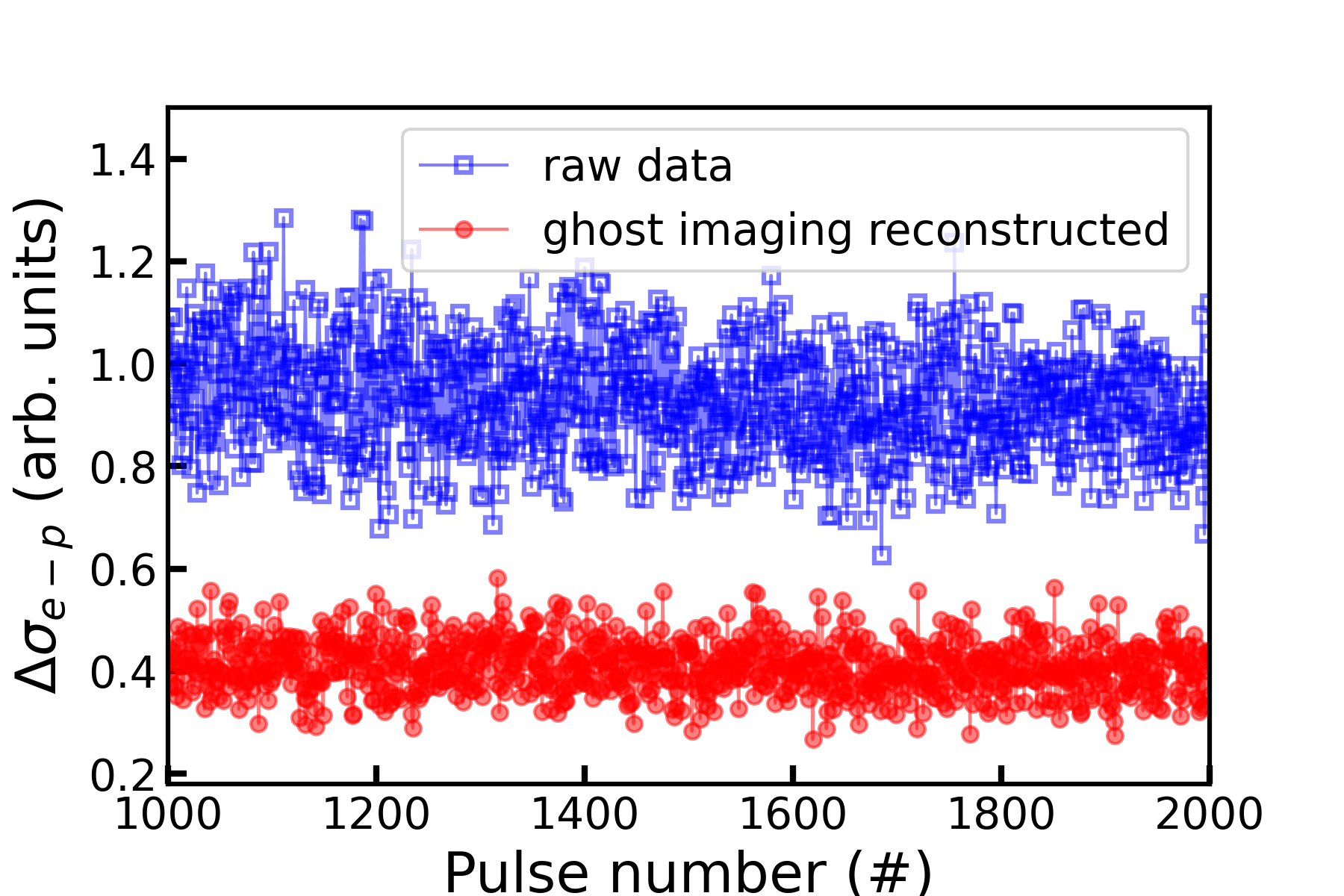}
\caption{\label{fig:std1} Statistical measure of improved spectral characterization using the ghost imaging algorithm. $\Delta\sigma_\mathrm{e-p}$ (Eqn. (\ref{std})), the standard deviation of the difference between electron-based and direct photon measurements, is plotted against pulse number. The $\Delta\sigma_\mathrm{e-p}$ are shown for before (blue, open square) and after (red, solid dot) the ghost-imaging reconstruction. The response matrix $R$ used here was determined using data from six electron time-of-flight spectrometers (eToFs) and all $N$ = 15337 shots. The results for 1000 shots are shown here.}
\end{figure}
 
 To quantify the performance of the reconstruction, we calculated the standard deviation of the difference signal between the electron-derived and the photon spectra $s_i$
 \begin{equation}
 \Delta\sigma_\mathrm{e-p}=\sqrt{\frac{\sum_{i=1}^{n}|(c_i-s_i)-(\bar{c}-\bar{s})|^2}{n}}
 \label{std}
 \end{equation}
 where $\bar{s}$ and $\bar{c}$ are the mean value of spectrometer and PES array measurement, respectively, $n$ is the number of spectrometer pixels. Depending on the situation, the value of $c_i$ is either interpolated PES array data of one eToF or the ghost imaging reconstructed spectrum. As shown in Fig.~\ref{fig:std1} the deviation $\Delta\sigma_\mathrm{e-p}$ of reconstructed spectrum drops to half of the original value, which indicates the improvement after reconstruction. In addition, the smaller fluctuation of the deviations means that the reconstructed spectrum is stable and more reliable than the raw electron-derived spectrum. The significantly better matching of the spectrum after ghost-imaging is further confirmed by a good correlation between reconstructed and photon spectrum i.e. averaging 0.72 Pearson correlation coefficient across the spectrum (see Supplementary Note 2).

\subsection*{Predictive power and performance analysis}
\begin{figure}
\centering
\includegraphics[width=0.5\textwidth]{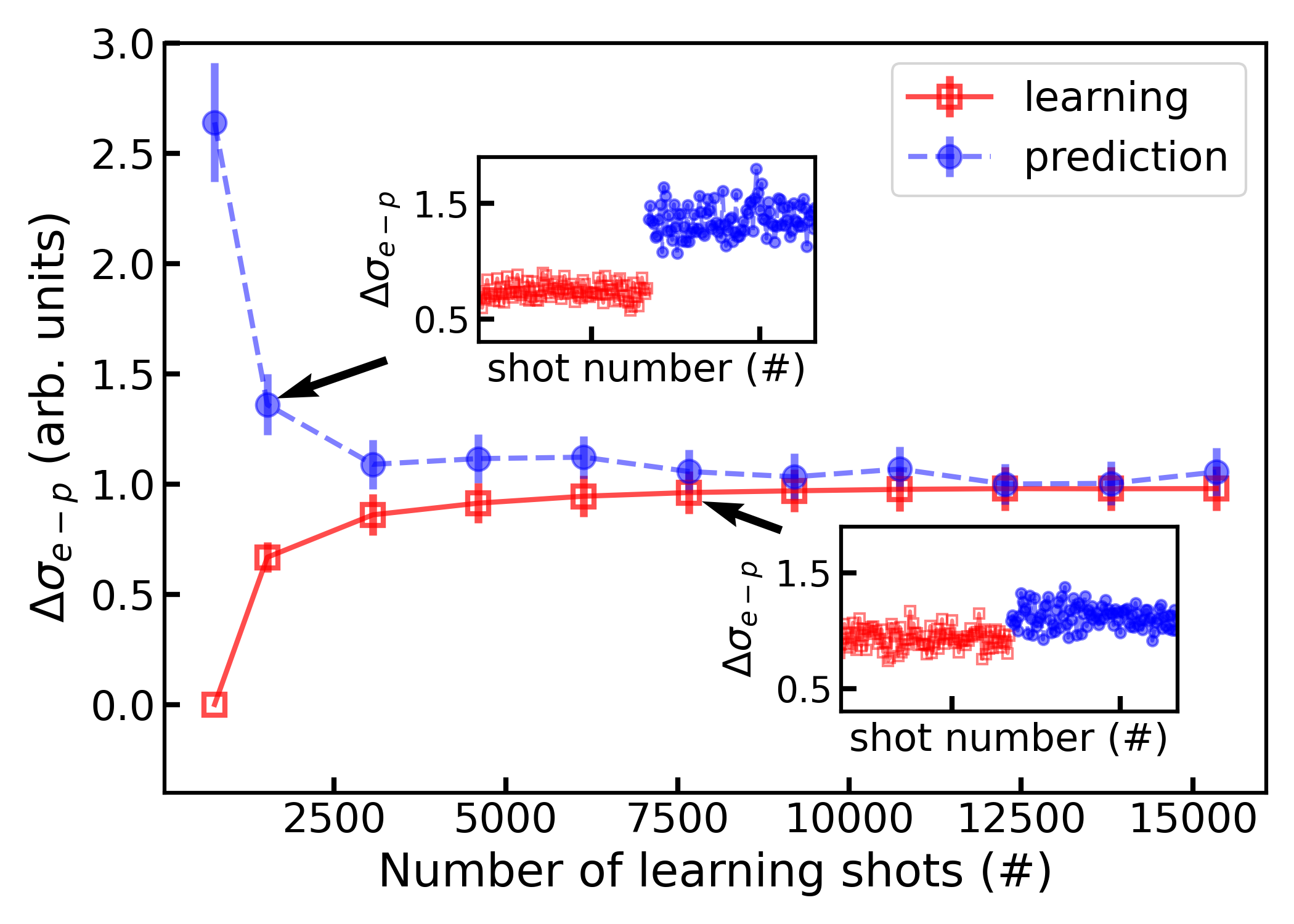}
  \caption{\label{fig:STD_num_shots} Predictive capability of the ghost imaging algorithm. $\Delta\sigma_\mathrm{e-p}$ (Eqn. (\ref{std})), the standard deviation between the ghost-imaging reconstructed and the true photon spectrum, is plotted as a function of the number of shots used in the regression to determine $R$. Prediction refers to $\Delta\sigma_\mathrm{e-p}$ for 100 shots not used in the learning regression.  The accuracy of the prediction increases (i.e. $\Delta\sigma_\mathrm{e-p}$ decreases) as the number of learning shots is increased. The error bars represent standard deviations of $\Delta\sigma_\mathrm{e-p}$. The inset at 1433 learning shots shows a marked difference between prediction and learning (0.6 vs 1.4), whereas the inset at 7568 learning shots shows a relatively small difference (0.9 vs 1.1). }
\end{figure}

One of the most interesting aspects of the ghost imaging method is its predictive power for future shots.  This requires numerous "learning" shots to obtain a converged response matrix.  Data from six eToFs were used and the response matrix learned from different number of shots is then used to predict the spectra for 100 new shots that were not used in the regression. As shown in Fig.~\ref{fig:STD_num_shots}, when fewer shots are used, the deviation of the learning shots is small because the regression is under-determined. Meanwhile, the deviation for the new shots is large, indicating a poor predictive power of an unconverged response matrix. As the information from more shots are included in the regression process, the deviation of learning shots rises, whereas the deviation for the new shots decreases indicating the gain of predictive power. The regression converges when the deviation of learning and prediction meet around 8000 shots (roughly 10 times the number of unknown variables i.e. PES array vector elements). It is important to note that as more shots are used the error bar for the prediction, which measures the fluctuation of the deviation, also decreases, which means the prediction becomes more stable. 

\begin{figure}
\centering
\includegraphics[width=0.5\linewidth]{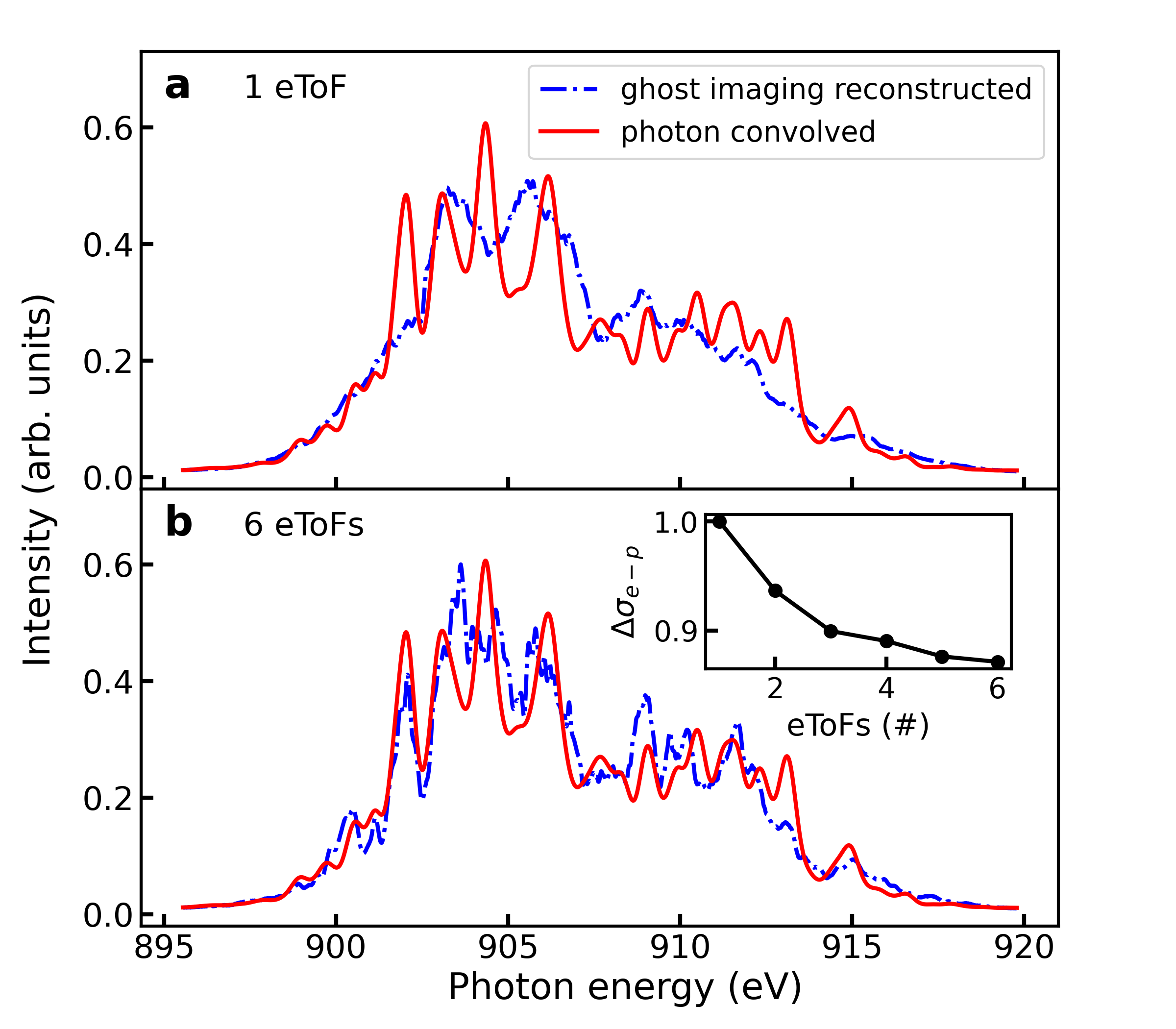}
\caption{\label{fig:single2} Improved single-shot reconstruction as a function number of electron time-of-flight spectrometers (eToFs) used. The reconstructed spectrum and Gaussian convolved spectrum using one eToF \textbf{(a)} and six eToFs \textbf{(b)}. All the shots are used to get a converged result. The inserted plot shows the decrease in  $\Delta\sigma_\mathrm{e-p}$ (Eqn. (\ref{std})) when more eToFs are used.}
\end{figure}

Combining electron spectra from several eToFs increases the signal intensities and suppresses the noise. However, the eToF signals cannot be added on top of each other directly, due to the different calibrations of each eTOF. As mentioned before, we put the signals from different eToFs together to form a larger vector and differences of eToFs are automatically taken into account when calculating the response matrix. The upper panel and the lower panel of Fig.~\ref{fig:single2} show the spectrum of a random shot where one eToF with strongest signal and six eToFs are used, respectively. Comparison with the Gaussian convolved spectrometer measurement clearly indicates the advantages of using six eToFs. The inset plot shows that the normalized deviation drops gradually from 1 for one eToF to 0.87 with six eToFs. The data from different eToFs complement with each other and improve the correlation with the spectrometer measurement thus resulting in the better overall reconstructed spectrum.

Our analysis indicates that the performance of the ghost imaging reconstruction depends on experimentally controllable parameters, number of shots used, number of eTOFs used. Ghost imaging is based on the correlation between the object measurement and the reference measurement, i.e. the sensitivity of the PES array signals to the fluctuations of the incident spectrum as measured by a grating spectrometer. Obtaining better correlation function and reconstruction, i.e. response matrix and spectrum with higher resolution, requires more data points within the Ne 1s photoelectron peak as well as high signal-to-noise ratio. More data points in the eTOF spectrum can be obtained by increasing the retardation voltages to slow the electrons, using larger drift length tubes, or, more simply by increasing the digitizer sampling rate which is presently 2 GHz.  In addition, increased detection sensitivity can be readily achieved by using more eToFs or by increasing the gas density to produce more photoelectrons and a higher signal-to-noise ratio (see discussion in Supplementary Note 3).


\section*{Discussion}

\begin{figure}[!htp]
\centering
\includegraphics[width=0.5\textwidth]{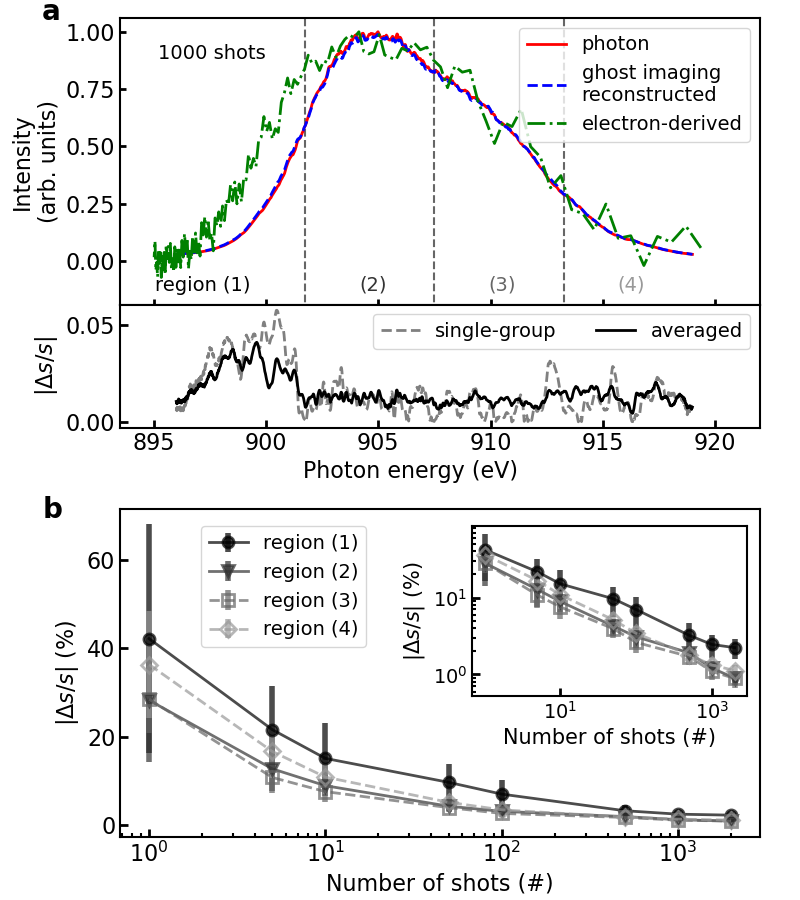}
\caption{\label{fig:averaged} Averaged ghost-imaging-reconstructed versus averaged photon spectra. (\textbf{a}) Upper panel shows the averaged spectrum for a group of 1000 shots for: photon (red solid line), ghost-imaging reconstructed (blue dashed line), and electron-derived, i.e. without application of ghost-imaging (green dot-dashed line). Note the electron-derived spectrum fails to reproduce the averaged photon spectrum in contrast to the ghost-imaging reconstruction. The spectrum is divided into regions 1-4, going from low to high photon energies. Lower panel shows corresponding spectral percentage deviation between the ghost-imaging reconstructed and true photon spectra for a single 1000-shot group (dashed) and averaged over seven 1000-shot groups (solid).  (\textbf{b})The normalized deviation $\Delta s/s$ (Eqn. (\ref{mean}))for each spectral region 1-4 as a function of number of averaged shots. The error bars represent the standard deviation of $\Delta s/s$.}
\end{figure}

Despite the considerable improvement to achieve resolution {$\Delta E/E \sim 1/2000$}, the ghost-imaging reconstruction of a single-shot XFEL spectrum under the current experimental conditions can not characterize completely the SASE structure containing inter-spike spacings down to $\sim 0.1$ eV, corresponding to a required resolution of {$\Delta E/E \sim 1/10000$}. This is simply due to the relatively sparse sampling rate (0.5 ns) and the relatively short flight path of the current eTOF spectrometer array \cite{laksman2019commissioning}. We note that there is a three-fold increased flight path in the PES array \cite{Walter:ys5104} at the Linac Coherent Light Source \cite{duris2020tunable}. The increased flight path will yield a corresponding increase in the number of eTOF sampling points and resulting resolution.  Compared to the current demonstration, a two-fold increased sampling rate combined with a two-fold increased time-of-flight yields four-fold increase in resolution - giving $\sim 120$meV FWHM for a resolving power of 8000. For this particular case using the Ne $1s$ photoelectron, the width of the final state Ne $1s^{-1}$ of 0.27 eV represents another barrier that ghost imaging can circumvent. 

For other applications, the ability to reconstruct an averaged spectral profile with high accuracy is of considerable interest.  High-precision incident spectra obtained by averaging are extremely useful for transient absorption measurements, e.g. when a dispersive spectrometer is placed after the sample one can extract spectral features below the SASE bandwidth, as was done previously in the XUV spectral range to observe strong-field induced modifications of the lineshape of a doubly-excited state in He \cite{Ott2019PRL}.  This averaging method is very common for pump/probe transient absorption experiments, e.g. with broadband soft x-ray high-harmonic generation (HHG) sources where single-shot spectra sequentially taken with pump-on and pump-off configurations are averaged to obtain spectral transients \cite{Attar-2017-Science,Pertot2017Science}. With the present transparent beamplitter, this can be extended to soft x-ray energies that are difficult to access with HHG.

The effect of shot averaging for ghost-imaging reconstruction is shown in Fig.~\ref{fig:averaged} where the deviation between the ghost-imaging-reconstructed spectrum and the grating spectrum is shown.  These deviations were evaluated as follows: measurements were randomly selected to form 7 groups of spectra with $X$ shots in each group.  The spectra within each group were then ensemble averaged. Fig.~\ref{fig:averaged}{\bf a} shows the result for $X$=1000 shots with the lower panel displaying the normalized deviations ($\Delta s/s$) for photon energies, $i$:
\begin{equation}
|\Delta s/s|_i=\frac{\sum_{k=1}^{X}\left| c_i^k-s_i^k \right| /s_i^k}{X}
\label{mean}
\end{equation}
where the upper index $k$ denotes the shot number. The deviation for one group is shown as a gray-dashed line, and that for the average of
7 groups is shown as a black-solid line.

The deviations show a spectral dependence as is clear from the lower panel of Fig.~\ref{fig:averaged}{\bf a} .  The data were divided into 4 regions separated at 901.75, 907.5, 913.25 eV. The  percentage deviations ($\Delta s/s$) for each region are shown in Fig.~\ref{fig:averaged}{\bf b} as a function of the number of shots averaged. The $\Delta s/s$ decreases from around 28\% for single-shot to 3\% and 1\% for 100 and 1000 shots, respectively. It is clear that the ghost imaging reconstructed spectrum matches significantly better with the grating spectrum after averaging hundreds of shots. This demonstrates the accuracy of the calculated response matrix.  If the single-shot deviation comes from random noise, the deviation after averaging $X$ shots will be proportional to $1/\sqrt{X}$.  The log-log plot inset in Fig. \ref{fig:averaged}{\bf b} clearly shows a negative power relationship. A curve fit of the data in region (2) yields a function of $1/X^{0.45}$, which is close to the expected $1/\sqrt{X}$.  We note that the deviation is somewhat larger in region (1) which corresponds to the low-photon-energy, low-photoelectron-kinetic-energy region. This may due to the small PES array signals (corresponding to a low signal-to-noise ratio) for electrons with lower kinetic energy. Region (1) also has a relatively small 0.53 Pearson coefficient (see further discussion in Supplementary Note 2). Averaging to achieve a $\sim1\%$ precision after $\sim1000$ shots (requiring only 1 ms at MHz repetition rates using the non-invasive photoelectron spectroscopy scheme) can foster the realization of transient absorption measurements at FELs \cite{Ott2019PRL}.

Returning to the single-shot spectral profile measurements, the ghost-imaging reconstruction demonstrated here is easy to apply for better calibration and resolution improvement of other instruments designed for XFEL diagnostics such as the newly inaugurated angle-resolved photoelectron spectroscopy (ARPES) instrument at LCLS \cite{Walter:ys5104}. After a training period using multiple eTOFs in conjunction with a high-resolution spectrograph to obtain a converged response matrix, spectral profiles of new shots with enhanced resolution can be obtained at MHz repetition rates.  The processing time to obtain the response matrix for the current dataset was $\sim$20 min using a single core and can be readily sped up for rapid implementation during an XFEL beamtime. From an XFEL-machine perspective, the enhanced energy resolution for an ARPES-based x-ray photon diagnostic, which has already demonstrated characterization of spatial, temporal and polarization properties, is a sought-after breakthrough since it enables a deeper understanding of the machine operation and allows for a fast-feedback on SASE-formation characteristics.  

From a scientific perspective, the high-resolution, non-invasive single-shot incident x-ray spectrum characterization described here represents a step forward for techniques that require incident spectral characterization such as s-TrueCARS \cite{cavaletto2021high} and others that take advantage of the intrinsic stochastic nature of XFEL pulses \cite{Kimberg2016SD,ratner2019pump,kayser2019core}.  Our ghost-imaging inspired gaseous beamsplitter will allow such experiments at high repetition rates (MHz) and also in the soft x-ray regime. For example, it will enable photon-in/photon-out transient multidimensional spectroscopy at XFELs to explore nonlinear effects arising from propagation through dense absorbing media \cite{Li2020PRA}.  Moreover, an incident pulse spectral characterization should enable a greater depth of understanding for processes where the SASE-pulse structure can play an important role, such as the recently-discovered transient resonance phenomena in core-hole dynamics in gaseous media which rely on mapping out narrow energy levels of highly elusive states of matter \cite{Mazza-2020-PRX}, SASE-FEL studies of chiral dynamics using photoelectron circular dichroism which have been compromised by averaging over subtle dynamics due to the large bandwidth \cite{Ilchen2021CommChem} and resonance-enhanced scattering for single-particle imaging \cite{Ho-2020-NatComm}. 

In summary, the use of a ghost-imaging algorithm with photoelectron spectrometer array as a transparent beamsplitter, as is currently available, can provide the incident SASE XFEL spectrum at high resolution on a single-shot basis and averaged SASE spectra at high precision.  The single-shot spectra, combined with polarization information from the eTOF array \cite{viefhaus2013variable,Walter:ys5104}, will allow users to extract complete two-dimensional spectra in a few thousand shots at high repetition rates and high intensities for studies of nonlinear, polarization-dependent x-ray phenomena using correlation techniques. The averaged spectra can be used to obtain high precision transient absorption spectra with resolution much higher than the SASE bandwidth. The combination of ghost-imaging with a photoelectron spectrometer array is expected to be a powerful asset for both the technological and scientific development of x-ray spectroscopies at XFELs. 


\section*{Methods}
\subsection*{XFEL photon delivery}
Our experiment was performed on the SQS (Small Quantum Systems) branch of the SASE3 beamline at the European XFEL \cite{decking2020mhz,tschentscher2017photon} with SASE soft x-rays at 910 eV central photon energy. The averaged FWHM bandwidth of the XFEL pulses was 9 eV and the standard deviation of pulse energy fluctuation was 3\% for an average pulse energy of 3.8 mJ as measured with an x-ray gas monitor detector (XGM). The XFEL ran at a 10 Hz repetition rate over a 25 minutes data acquisition time to obtain $N =15337$ shots.  

\subsection*{Electron measurement -  photoelectron spectrometer array}
The PES array is located far from any x-ray focus points and the SASE beam spot size was estimated to be $\sim5$ mm in diameter, ensuring the photoionization remains within the linear regime. The number of photoelectrons generated is proportional to the product of the gas density and the photoionization cross section. The base background gas pressure in the chamber is $1\times 10^{-8}$ mbar and the pressure of gas injected was adjustable from $1\times 10^{-7}$ mbar to $1\times 10^{-5}$ mbar. The gas density was set to $2.5\times 10^{-7}$ during our experiment. The retardation voltage of 30 V slowed photoelectrons with 40 eV initial kinetic energy to 10 eV. The total time-of-flight for Ne $1s$ photoelectrons from the interaction region to the detector is $\sim 60$ ns (see ToF signals in Supplementary Note 1). The detector signals from the microchannel plate (MCP) stack were recorded every 0.5 ns. The electrons were slowed in order to create a $1s$ photoelectron peak with more ToF sampling points while keeping the signal well above the background. 

\subsection*{Photon measurement - grating spectrograph}
The FEL beam (3.8 mJ/pulse) was attenuated prior to the spherical mirror with Kr gas (transmission = 35.5\%); combined with the grating efficiency of 36\%, 0.49 mJ was incident on the screen.  The VLS grating has a groove density of 150 lines/mm, length of 120 mm and incidence angle of 12.3 mrad, with the imaging screen located at 99 m distance from the grating, in the focus of the spherical premirror. The present VLS grating intercepted the central portion of the x-ray beam, which for the spatially unchirped beam expected here under standard 
XFEL operating conditions, is representative of the energy spectrum of the full beam.  Slitting options upstream of the PES array are available to account for special chirped XFEL operating modes. The emitted fluorescence from the Ce:YAG screen was detected by a camera to produce the photon spectrum \cite{serkez2020opportunities}. The spectral range recorded on the YAG screen was from 895.5 eV to 919.8 eV and was spread over 1900 pixels.  The estimated resolving power for the spectrometer, based on independent measurements, is $E/\Delta E = 10000$ \cite{Gerasimova_prep}. This allows to resolve the single SASE spikes of the XFEL pulses, which show a minimum spacing of $\sigma=90$ meV in the present measurements.

\subsection*{Data analysis}
The data analysis was carried out on the European XFEL Maxwell server using Jupyter notebook with a single core. The time consumed to calculate the response matrix of the PES array depends on the number of eTOF data included and the number of shots used. To obtain fully converged results, the number of shots used should be around 10 times the number of elements in the PES array vector $c$. The calculation takes 30 seconds for 1 eTOF (137 points) using 1200 shots, and 20 minutes for 6 eTOF (822 points) using 8000 shots. It is straightforward to use several cores with  MPI to shorten the computing time significantly. Note that no filters are applied during the analysis. Filters keeping the shots with good correlations between PES array and spectrometer integral signals were tried. However, it failed to improve the performance of ghost-imaging method, which benefits from the variance of SASE pulses.

\section*{Data Availability}
The experimental data were collected during beamtime 2935 at the European XFEL.  The metadata are available at [https://in.xfel.eu/metadata/doi/10.22003/XFEL.EU-DATA-002935-00].

\section*{References}
\bibliographystyle{naturemag_noURL} 
\bibliography{mybib} 

\section*{Acknowledgments}
This work was supported by the U.S. Department of Energy, Office of Science, Basic Energy Science, Chemical Sciences, Geosciences and Biosciences Division under contract number DE-AC02-06CH11357.  We thank Chuck Kurtz for assistance with Figure 1. We thank Natalia Gerasimova for assistance with the grating spectrometer setup and operation.  We acknowledge European XFEL in Schenefeld, Germany for provision of the x-ray free-electron laser beam time at the SASE 3 undulator and thank the staff for their assistance.

\section*{Author Contributions}

K.L. and L.Y. conceived the use of ghost-imaging reconstruction for high-resolution spectral characterization with multiple photoelectron spectrometers.  T.M., J.L., M.M., K.L., L.Y. planned and designed the experiment. T.M. and J.L. set up the experimental configuration and performed data collection together with K.L., D.K., G.D., A.P., M.M., M.I. and L.Y.. S.S. and N.R. provided discussions on spectral averaging analysis.  K.L. performed data analysis and together with L.Y. interpretation. K.L. and L.Y. wrote the paper. All authors discussed the results and contributed to the final manuscript.

\section*{Competing Interests}
The authors declare no competing interests.


\end{document}


\title{Supplementary information of "Ghost-imaging-enhanced non-invasive spectral characterization of stochastic x-ray free-electron-laser pulses"} 

\author{Kai Li}\email{kail@anl.gov}
 \affiliation{Department of Physics, The University of Chicago, Chicago, IL 60637 USA}
 \affiliation{Chemical Sciences and Engineering Division, Argonne National Laboratory, Lemont, IL 60439 USA}
  

\author{Joakim Laksman}
\affiliation{European XFEL, Holzkoppel 4, 22869 Schenefeld 
Germany}

\author{Tommaso Mazza}
\affiliation{European XFEL, Holzkoppel 4, 22869 Schenefeld 
Germany}

\author{Gilles Doumy}
\affiliation{Chemical Sciences and Engineering Division, Argonne National Laboratory, Lemont, IL 60439 USA}

\author{Dimitris Koulentianos}
\affiliation{Chemical Sciences and Engineering Division, Argonne National Laboratory, Lemont, IL 60439 USA}

\author{Alessandra Picchiotti}
\affiliation{The Hamburg Centre for Ultrafast Imaging, Hamburg University, Luruper Chaussee 149, 22761, Hamburg, Germany}

\author{Svitovar Serkez}
\affiliation{European XFEL, Holzkoppel 4, 22869 Schenefeld 
Germany}

\author{Nina Rohringer}
\affiliation{Center for Free-Electron Laser Science CFEL, Deutsches Elektronen-Synchrotron DESY, Notkestra{\ss}e 85, 22607 Hamburg, Germany}
\affiliation{Department of Physics, Universität Hamburg, 20355 Hamburg, Germany}

\author{Markus Ilchen}
\affiliation{Deutsches Elektronen-Synchrotron DESY, Notkestra{\ss}e 85, 22607 Hamburg, Germany}

\author {Michael Meyer}
\affiliation{European XFEL, Holzkoppel 4, 22869 Schenefeld 
Germany}

\author{Linda Young}
 \email{young@anl.gov}
 \affiliation{Department of Physics, The University of Chicago, Chicago, IL 60637 USA}
 \affiliation{Chemical Sciences and Engineering Division, Argonne National Laboratory, Lemont, IL 60439 USA}
 \affiliation{James Franck Institute, The University of Chicago, Chicago, IL 60637 USA}

\date{\today}

\maketitle

\date{September 2021}

\clearpage
\section{Supplementary Note 1: details of eToF performance}

The single-shot and shot-averaged signals ($y$) from one eToF along the polarization direction are shown in Supplementary Fig.~\ref{fig:tof}. The standard deviation of the signals $\sigma_y(\omega)$ at different photon energies are shown in shaded region. The photoionization of different atomic sub-shell electrons and Auger decay lead to different peaks as denoted in the figure. With $2.4\times 10^{-7}$ mbar and 0.3 Mb Ne 1s photoionization cross section at 910 eV \cite{Coreno1999PRL}, an incident x-ray pulse with an average of $2.6\times 10^{13}$ photons would generate around 144000 photoelectrons per pulse. Given the geometry of the PES array \cite{buck2012online}, $\sim 300$ Ne 1s photoelectrons are detected by each eToF along the polarization direction. The x-ray beam diameter in the PES array is estimated to be 5 mm. For a 40 eV Ne $1s$ photoelectron, this beam diameter creates a flight path ToF delay of $\Delta t=1.3$ ns, which limits the raw PES resolution to above 0.9 eV. The $\sigma_y(\omega)$ represents the shot-to-shot fluctuation of SASE pulses plus the noise of the ToF. As shown below in Supplementary Fig.~\ref{fig:pressure}\textbf{a}, the standard deviation to mean value ratio ($\sigma_y(\omega)/\bar{y}(\omega)$) is calculated to represent the noise-to-signal ratio of the eToF signals. The background of the eToF signals is assumed to be equal to the average signal between 100 to 120 ns. It is subtracted on a shot-by-shot basis to obtain background-free signals.

\renewcommand{\figurename}{Supplementary Figure}

\begin{figure}[!htp]
\includegraphics[width=0.85\linewidth]{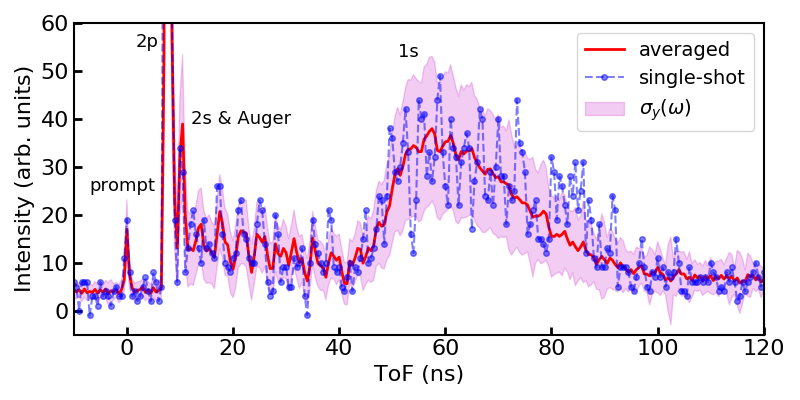}
\caption{\label{fig:tof} Photoelectron signal of one eToF along the polarization direction. The averaged and single-shot signal is in red solid and blue dashed line, respectively. The
standard deviation of signals is in pink shaded area.}
\end{figure}

\section{Supplementary Note 2: correlation analysis}
The analysis of correlations between the true photon spectrum and ghost-imaging-reconstructed spectrum are shown in Supplementary Fig.~\ref{fig:correlation}. The spectral intensity is integrated over 1-eV energy bins. For the bin between 908 to 909 eV, the correlation of 5000 shots is shown in (\textbf{a}). A good correlation is indicated by the dots along the line (gray dashed) with a slope equaling one. The Pearson correlation coefficient \cite{moore2013basic} across the spectrum is shown in (\textbf{b}). The Pearson coefficient stays around 0.75 at the central photon energy and decreases to 0.5 at the two wings.

\begin{figure}[!htp]
\includegraphics[width=0.9\linewidth]{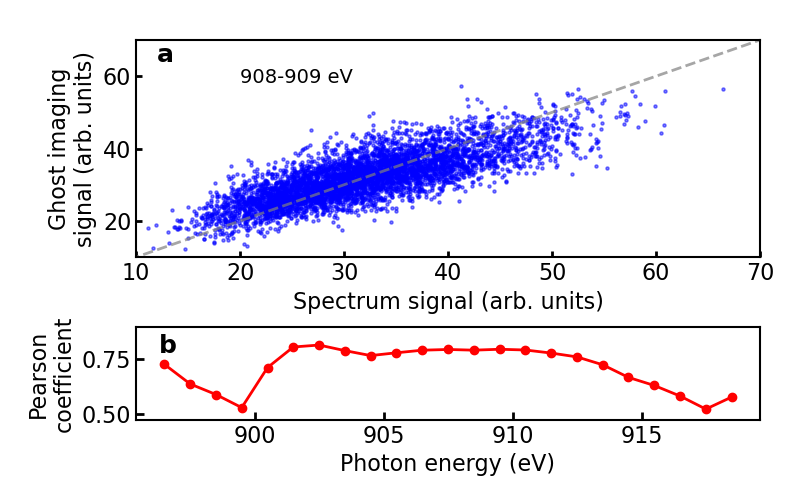}
\caption{\label{fig:correlation} The correlation between the true photon spectrum from the spectrometer and the ghost-imaging-reconstructed spectrum. \textbf{a} The correlation between photon spectrum and ghost imaging reconstructed spectrum integrated within 908 - 909 eV. \textbf{b} Pearson correlation coefficient at different photon energies.}
\end{figure}

\section{Supplementary Note 3: limitations to single-shot reconstruction}
The ghost imaging reconstruction works by learning the response of PES array and then using the response matrix to predict the incident spectrum based on the PES array signal. The predictive power is limited by the information recorded in the PES array measurement, i.e. the PES array signal-to-noise ratio and the number of ToF points within the Ne 1s photoelectron peak. 

\begin{figure}[!htp]
\includegraphics[width=0.95\linewidth]{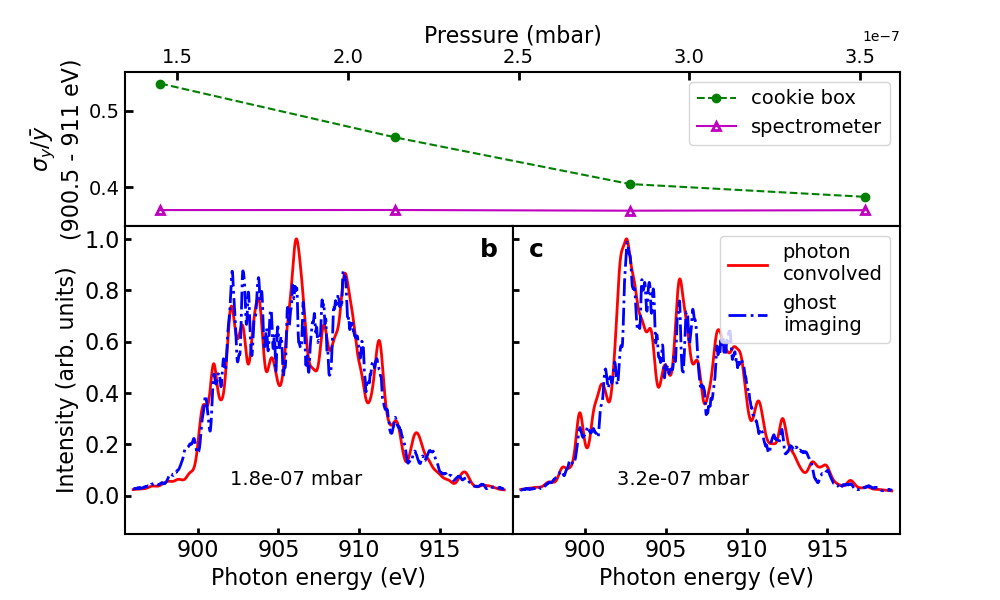}
\caption{\label{fig:pressure} Improved signal to noise with increased Ne pressure. (\textbf{a}) Standard deviation to mean value ratio of PES array and spectrometer signal with different gas pressure. The ghost imaging reconstructed spectrum and convolved photon spectrum at (\textbf{b}) $1.8\times 10^{-7}$ mbar, (\textbf{c}) $3.2\times 10^{-7}$ mbar. }
\end{figure}

To estimate the noise level of signals, we calculated the standard deviation to mean value ratio at different photon energies $\sigma_y(\omega)/\bar{y}(\omega)$ for the PES array and spectrometer measurements. Supplementary Fig.~\ref{fig:pressure}\textbf{a} shows the mean value of $\sigma_y(\omega)/\bar{y}(\omega)$ within the main peak $900.5 \leq \omega \leq 911$ eV with different gas densities inside the PES array. The averaged standard deviation to mean value ratio (SMR) of spectrometer stays the same while the SMR value of PES array decreases with higher gas pressure. As mentioned before, the standard deviation of PES array signals comes from both the fluctuation of SASE pulse and the noise in the eToF signal. Assuming the noise of spectrometer can be ignored, the deviation of SMR value from the value of spectrometer measures the noise level of PES array. As expected, with higher gas density, the SMR of PES array decreases and approaches that of the spectrometer, which indicates lower noise level, i.e. better signal-to-noise ratio, when more photoelectrons are generated. The ghost imaging reconstructed spectrum and convolved photon spectrum of a random shot at different gas pressure $1.8\times 10^{-7}$ and $3.2\times 10^{-7}$ mbar are shown in Supplementary Fig.~\ref{fig:pressure}\textbf{b} and \textbf{c}, respectively.

\begin{figure}[!htp]
\includegraphics[width=0.9\linewidth]{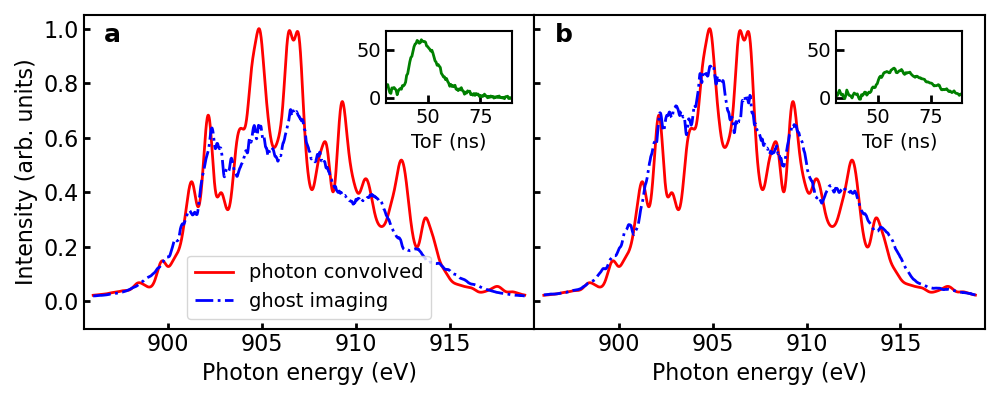}
\caption{\label{fig:tofs} The ghost imaging reconstructed single-shot spectrum and convolved photon spectrum with ToF peak at $\sim$ 50 ns (\textbf{a}) and $\sim$ 60 ns (\textbf{b})}
\end{figure}

The ghost imaging reconstructed spectrum and convolved photon spectrum with ToF peaks at $\sim$ 50 and $\sim$ 60 ns are shown in Supplementary Fig.~\ref{fig:tofs}\textbf{a} and \textbf{b} respectively. The inserted figures show the ToF signals. There are around 40 ToF points within the major part of Ne 1s peak located $\sim$ 50 ns and 20 more points within the major peak at $\sim$ 60 ns. Clearly the measurement with more ToF points contains more SASE information, thus gives a better reconstructed spectrum. Note that there is a trade-off between keeping a good signal-to-noise ratio and slowing down electrons to obtain more ToF points. The signals becomes smaller with the electrons spread out over a larger ToF range. The pressure used in the experiment was below the maximum allowed in the chamber $1\times 10^{-5}$ mbar \cite{laksman2019commissioning}.  

\bibliographystyle{naturemag_noURL} 
\bibliography{SI_bib} 
